\documentclass[a4paper,11pt]{article}
\pdfoutput=1 % if your are submitting a pdflatex (i.e. if you have % images in pdf, png or jpg format)

\usepackage{jheppub} % for details on the use of the package, please see the JHEP-author-manual
\usepackage[T1]{fontenc} % if needed
\usepackage[all]{hypcap}
\usepackage{slashed} % for Feynman slash notation
\usepackage{multirow} % more control on table columns/rows
\usepackage{booktabs} % for better looking tables
\usepackage{parskip} % nice white line spacing between paragraphs
\usepackage[section]{placeins} % Prevents floats moving to another section
\usepackage{pifont} % checkmarks and crosses
\usepackage{mathtools} % for dcases environment

\graphicspath{ {plots/} }

\newcommand{\subh}{\text{\tiny H}}
\newcommand{\subt}{\text{\tiny T}}
\newcommand{\subw}{\text{\tiny W}}
\newcommand{\subz}{\text{\tiny Z}}
\newcommand{\subl}{\text{\tiny L}}
\newcommand{\subr}{\text{\tiny R}}

\title{\boldmath Top Partner Discovery in the $T\to tZ$ channel at the
  LHC} 
\author{J\"urgen Reuter and}
\author{Marco Tonini}
\affiliation{DESY Theory Group\\Notkestr. 85, D-22607 Hamburg, Germany}
\emailAdd{juergen.reuter@desy.de}
\emailAdd{marco.tonini@desy.de}
\preprint{\small DESY 14-141}
\arxivnumber{1409.6962}

\abstract{
In this paper we study the discovery potential of the LHC run II
  for heavy vector-like top quarks in the decay channel to a top and a
  $Z$ boson. Despite the usually smaller branching ratio compared to 
  charged-current decays, this channel is rather clean and allows for a
  complete mass reconstruction of the heavy top. The latter is
  achieved in the leptonic decay channel of the $Z$ boson 
  and in the fully hadronic top channel using boosted jet and jet
  substructure techniques. To be as model-independent as possible, a 
  simplified model approach with only two free parameters has been
  applied. The results are presented in terms of parameter space
  regions for $3\sigma$ evidence or $5\sigma$ discovery for such new
  states in that channel.
}

\begin{document}

%%%--- Title Page ---%%%
\maketitle
\flushbottom

%%%--- Introduction ---%%%
\section{Introduction}
\label{sec:intro}
With the advent of the Large Hadron Collider (LHC), a whole new range
of energies is opening up for experimental particle physics, namely
the range from the electroweak scale $v$ up to the multi-TeV
regime. Within the first 2010-2012 run of the LHC crucial results
have been already collected, most notably the discovery of a (light)
Higgs boson with mass $m_{h} \sim 125 \, \textrm{GeV}$, publicly
announced on the 4th of July 2012
\cite{Chatrchyan:2012ufa,Aad:2012tfa}. Also remarkable are the
(preliminary) measurements of the Higgs couplings and production
modes, which are turning out to be as predicted by the Standard Model:
no significant sign of new phenomena has been observed so far. This is
starting to provide severe constraints on possible theories that
differ significantly from the Standard Model at the probed energies. 

Despite this enormous success, we know that the Standard Model cannot
describe all phenomena we have observed so far. In particular, the
absence of a possible candidate to describe the Dark Matter and Dark
Energy hinted by various cosmological and
astrophysical observations as well as the missing CP violation for the
explanation of the baryon-antibaryon asymmetry represent the main
experimental results that cannot be accommodated within the SM.

Furthermore, different theoretical motivations are considered as
issues of the actual Standard Model formulation above the electroweak
scale. The most notable one is the fine-tuning problem: a light
(fundamental) Higgs boson implies large accidental cancellations
between different and in principle uncorrelated physical quantities,
due to its large radiative sensitivity to possible higher scales in
the theory. In a ``natural'' theory, large cancellations among
uncorrelated terms should either not be present, or explained by means
of symmetry arguments. 

The issue of a necessary fine-tuning to account for a light Higgs
boson has always been the main guideline for possible model building
of Beyond the Standard Model (BSM) Physics: suitable new phenomena
should appear around the TeV energy scale in order to suppress the
large radiative corrections to the Higgs mass. The most sought-after
solution of the fine-tuning problem at the LHC is Supersymmetry
(SUSY). An alternative solution is given by strongly-coupled
extensions of the Standard Model. In this class of models, a new
strong interaction sector in assumed at some energy above the
electroweak scale, making the Higgs a composite object below the
compositeness scale. Since it does not make sense to speak of an
elementary scalar Higgs boson above the compositeness scale, at low
energies the Higgs mass is thus at most sensitive to the value of the
compositeness scale. In this sense, assuming a strong sector as
UV-completion of the Standard Model prevents dangerous fine-tuning
requirements to account for the observed Higgs mass. However, in a
generic strongly interacting extension of the Standard Model, the
compositeness scale would be close to the Higgs mass, causing a
conflict with electroweak precision observables and direct searches
for heavy resonances.  

A consistent way to implement a strongly coupled UV-completion of the
Standard Model has led to models in which the
Higgs arises as pseudo-Goldstone boson of some spontaneously broken
global symmetry of the strong sector at a scale $f \gg v$. The Higgs
boson can thus be much lighter than other possible states of the
composite sector, in complete analogy with the low-energy QCD
description, where the pions arise as a set of scalar states naturally
lighter than the compositeness scale $\Lambda_{\text{QCD}}$, with all
other resonances at higher masses. These models are generically called
Composite Higgs models. 

In particular, light partners of the SM top are a key ingredient for
the naturalness argument of different BSM models, in order to cut off
the quadratic UV-sensitivity of the Higgs mass squared parameter from
SM top loops. This is a common feature for generic Supersymmetric and
Composite Higgs models. The main difference between supersymmetric top
partners (\emph{stops}) and top partners arising in strongly coupled
models is their different spin, spin 0 vs. spin 1/2,
respectively. The fermionic top partners are usually vector-like
particles.  

Contrary to sequential fourth-generation quarks, which are heavily
constrained already from Higgs boson searches, since they would yield
a large impact e.g.~in the one-loop induced processes like gluon
fusion production and diphoton decay of the Higgs, indirect bounds on
vector-like quarks are much weaker. Their effect on the Higgs
observables is indeed less dramatic than fourth generation quarks as
their vector-like nature allows to obtain a large Dirac mass 
without introducing a large Yukawa coupling to the Higgs. 

Both the ATLAS and CMS collaborations have recently performed
dedicated searches for top partners
\cite{ATLAS:2014:036,Aad:2014efa,TheATLAScollaboration:2013sha,CMS:2014fya,Chatrchyan:2013uxa,CMS:2014rda,CMS:2014aka,CMS:2013una}. Depending
on the particular branching ratio under investigation, the actual
limits on the top partner mass, at $\sqrt{s} = 8 \, \textrm{TeV}$ and
with up to $20 \, \textrm{fb}^{-1}$ of integrated luminosity, do not
exceed $700-800 \, \textrm{GeV}$. Most of these experimental searches
assume the new heavy quarks to be pair produced: however, searches
combining pair production with single production through electroweak
interactions will become an important feature in the future. Present
limits from the LHC start to enter the region in which single
production becomes comparable to pair production due to the smaller
phase space suppression, even if an electroweak coupling is involved. 

Many different theoretical analyses involving top partners have been
recently proposed, some of them exploiting tagging techniques
\cite{Contino:2008hi,AguilarSaavedra:2009es,Mrazek:2009yu,Gopalakrishna:2011ef,Vignaroli:2012nf,DeSimone:2012fs,Kearney:2013oia,Gopalakrishna:2013hua,Li:2013xba,Azatov:2013hya,Beauceron:2014ila,Ortiz:2014iza,Han:2014qia,Brooijmans:2014eja,Yang:2014usa,Endo:2014bsa,Gripaios:2014pqa,Matsedonskyi:2014mna,Backovic:2014uma,Basso:2014apa}. However,
a closer look to these references reveals that the top partner decay
$T \to Z \, t$ has not been thoroughly explored yet, because
it appears rather difficult at first glance. In particular, the
all-hadronic final state suffers from huge SM backgrounds, making the
alternative $T \to W \, b$ channel more suited for
all-hadronic analyses due to the enhanced branching ratio and the
possibility to exploit b-tagging. Furthermore, the channel involving
a leptonic decay of the $Z$ entails a large suppression from the $Z$
leptonic branching ratio, $\textrm{BR}(Z \to \ell^{+}
\ell^{-}) \sim 0.067$ ($\ell \equiv e, \, \mu$). A study of the $T \to Z \, t$ ``trilepton'' channel 
with both leptonic decays of the $Z$ boson and top quark has 
been first proposed in \cite{Brooijmans:2014eja} and recently published 
in \cite{Basso:2014apa}.

In order to test the nature of the top partner, it is important to
develop search strategies which might cover all possible channels,
especially for the foreseen LHC energy upgrade to $13 \,
\textrm{TeV}$. For this reason, we develop a search strategy
tailored for a charge-2/3 top partner optimised for its decay channel
$T \to t \, Z \to \left( q \, q^{\prime} \, b \right)
\, \left( \ell^{+} \ell^{-} \right)$, at the LHC with center-of-mass
energy of $\sqrt{s} = 13 \, \textrm{TeV}$ and integrated luminosity of
$300 \, \textrm{fb}^{-1}$. We present, with minimal assumptions on the
underlying model, a method to discover a possible top partner
signature with large statistical significance. More importantly, we
aim at a precise measurement of its invariant mass. 

Recently, ATLAS presented a $\sqrt{s} = 8 \, \textrm{TeV}$
search~\cite{ATLAS:2014:036,Aad:2014efa} optimised for either 
pair or single production of a top partner, subsequently decaying as
$T \to Z \, t$ with leptonic decay of the $Z$ boson. This
encouraged us to further analyse this rather unexplored process, in
order to provide an effective search strategy for the forthcoming $13
\, \textrm{TeV}$ LHC runs. 

The structure of the paper is the following. In section
\ref{sec:toptaggingintro} we briefly review different examples
of models comprising top partners in the context of strongly coupled
UV-completions of the SM. This is followed by a discussion of a
simplified-model approach for the simulation of top partner
signal events, and some details about top-tagging techniques useful
to tag the boosted regime of the top partner decay products. Section
\ref{sec:ttagsetup} presents the setup of our proposed analysis,
namely the event generation procedure, the reconstruction of physics
objects, and the definition of the dedicated selection cuts. Finally,
a thorough discussion of the results is presented in section
\ref{sec:ttagresults}, together with concluding remarks in section
\ref{sec:conclusion}.

%%%-- Section: Top partners and top tagging --%%%
\section{Top partners and top tagging}
\label{sec:toptaggingintro}

%-- Subsection: Models comprising top partners
\subsection{Models comprising top partners}
All differences on the underlying top-partner model depend on the
choice of the representation of the new quarks and on the assignment
of the quantum numbers. We will briefly discuss some examples of
top partners in the context of strongly coupled UV-completions of the
SM. 

A prominent class of models predicting light spin-1/2 vector-like
top partners is the class of Composite Higgs models
\cite{Kaplan:1983fs,Kaplan:1983sm,Kaplan:1991dc,Contino:2006nn,Agashe:2004rs,Contino:2006qr,Carena:2007ua,DeSimone:2012fs}. In
the minimal Composite Higgs scenario, the coset structure is
$SO(5)/SO(4)$. The main guiding principle is that the decays and
single production of the new partners are generated via mixing with
the standard quarks, induced by Yukawa interactions with the Higgs. In
particular, only the right-handed SM top quark $t_{\subr}$ is
promoted to a fully composite state belonging to a complete multiplet
(singlet) of the unbroken $SO(4)$ group, while the (elementary)
left-handed SM doublet $q_{\subl}$ is assumed to be embedded into an
incomplete $SO(5)$ multiplet and to couple linearly to the strong
sector.

The vector-like top partners are introduced as composite
bound states belonging to a complete multiplet $\Psi$ of the unbroken
group $SO(4)$: two cases are usually considered, namely $\Psi \sim
\mathbf{4}$ or $\Psi \sim \mathbf{1}$ under $SO(4)$. We will refer to
these two implementations as $M4_{5}$ and $M1_{5}$, respectively. In
the $M4_{5}$ case, the multiplet $\Psi$ includes two charge-2/3 top
partners $X_{2/3}, \, T$, one exotic charge-5/3 top partner
$X_{5/3}$, and a charge-1/3 bottom partner $B$: under the SM gauge
group, the four components of $\Psi$ decompose into two SM doublets
($T,\, B$) and ($X_{5/3}, \, X_{2/3}$) of hypercharge 1/6 and 7/6, 
respectively. In the  $M1_{5}$ case, only one $SU(2)$-singlet
charge-2/3 top partner $\tilde{T}$ is introduced.  

Assuming an embedding of the elementary SM doublet $q_{\subl}$ into an
incomplete fundamental representation $Q^{\mathbf{5}}_{L} \sim
\mathbf{5}$ of $SO(5)$, the following interactions involving the top
partners can be written down \cite{DeSimone:2012fs}: 
\begin{align}
  \mathcal{L}^{M4_{5}} &\supset i \, c_{1} \left(
    \bar{\Psi}_{\subr} \right)_{i} \gamma^{\mu} d_{\mu}^{i} \,
  t_{\subr} + y \, f \left( \bar{Q}^{\mathbf{5}}_{\subl}
  \right)^{I} U_{I \, i} \, \Psi_{\subr}^{i} + y \, c_{2} \, f
  \left( \bar{Q}^{\mathbf{5}}_{\subl} \right)^{I} U_{I \, 5} \,
  t_{\subr} + \, \text{h.c.}  
  \label{lagrM45} \\
  \mathcal{L}^{M1_{5}} &\supset y \, f \left(
    \bar{Q}^{\mathbf{5}}_{\subl} \right)^{I} U_{I \, 5} \,
  \Psi_{\subr} + y \, c_{2} \, f \left(
    \bar{Q}^{\mathbf{5}}_{\subl} \right)^{I} U_{I \, 5} \,
  t_{\subr} + \, \text{h.c.} 
  \label{lagrM15}
\end{align}
In particular, $d_{\mu}$ is the connection symbol defined in the CCWZ formalism
\cite{Coleman:1969sm,Callan:1969sn}, $U$ is the $5 \times 5$ Goldstone
boson matrix, $y$ is a Yukawa coupling controlling the mixing between
the composite and elementary states, $c_{1}, \, c_{2}$ are
$\mathcal{O}(1)$ parameters associated with the interactions of
$t_{\subr}$, and $f$ is the usual symmetry breaking scale of the
strong sector. For the model $M1_{5}$, a direct coupling of $\Psi$
with $t_{\subr}$ like the first term in eq.~\eqref{lagrM45} can be
removed with a field redefinition. Note that the operators
proportional to $y$ explicitly break the $SO(5)$ symmetry, since
$q_{\subl}$ is embedded into an incomplete $SO(5)$ multiplet, giving
rise to the leading contribution to the Higgs potential triggering the
electroweak symmetry breaking.  

It turns out that the couplings of the top partners to the Goldstone
bosons ($\phi^{\pm}, \, \phi^{0}$), which in the high energy limit
correspond to the longitudinal components of the gauge bosons
(Equivalence Theorem), and to the Higgs $h$, are proportional to
linear combinations of the couplings $y, \, c$ \cite{DeSimone:2012fs}: 
\begin{align}
  M4_{5}: &
  \begin{dcases}
    \ \phi^{+} \, \bar{X}_{5/3 \, \subl} \, t_{\subr} & :
    \, \sqrt{2} \, c_{1} \, g_{\Psi} \\ 
    \ \left( h + i \phi^{0} \right) \bar{X}_{2/3 \, \subl}
    \, t_{\subr} & : \, c_{1} \, g_{\Psi} \\ 
    \ \left( h - i \phi^{0} \right) \bar{T}_{\subl} \,
    t_{\subr} & : \, -c_{1} \sqrt{y^{2} + g_{\Psi}^{2}} +
    \frac{c_{2} \, y^{2}}{\sqrt{2} \sqrt{y^{2} +
        g_{\Psi}^{2}}} \\ 
    \ \phi^{-} \bar{B}_{\subl} \, t_{\subr} & : \, c_{1}
    \sqrt{2} \sqrt{y^{2} + g_{\Psi}^{2}} - \frac{c_{2} \,
      y^{2}}{\sqrt{y^{2} + g_{\Psi}^{2}}} 
  \end{dcases} 
  \label{M45GBcouplings} \\
  M1_{5}: &
  \begin{dcases}
    \ \left( h + i \phi^{0} \right)
    \bar{\tilde{T}}_{\subr} \, t_{\subl} & : \,
    \frac{y}{\sqrt{2}} \\ 
    \ \phi^{+} \bar{\tilde{T}}_{\subr} \, b_{\subl} & : \,
    y \, , 
  \end{dcases}
  \label{M15GBcouplings}
\end{align}
where $g_{\Psi} = M_{\Psi}/f$, $M_{\Psi}$ being the Dirac mass of the
top partner multiplet.  

These couplings govern the associated production of the different top
partners. In particular we see that the $SU(2)$-singlet top partner
$\tilde{T}$ can be copiously produced in association with a b-quark:
from eq.~\eqref{M15GBcouplings}, its coupling to the $W$ boson is
given by  
\begin{equation}
  \left( \frac{m_{\subw}}{M_{\tilde{\subt}}} \right) \cdot
  \text{coeff} ( \phi^{+} \, \bar{\tilde{T}}_{\subr} \,
  b_{\subl} ) = \left( \frac{m_{\subw}}{M_{\tilde{\subt}}}
  \right) \, y \equiv \frac{g \, g^{\ast}}{\sqrt{2}} \, , 
  \label{M15associatedcoupling}
\end{equation}
with $y$ of order $\mathcal{O}(1)$ to reproduce the SM top mass.

Furthermore, we can easily read off from eq.~\eqref{M45GBcouplings}
and \eqref{M15GBcouplings} the different branching ratios of all top
partners. For example, in the decoupling limit of $m_{\Psi}
\to \infty$, the branching ratios of the $M1_{5}$
$SU(2)$-singlet top partner $\tilde{T}$ are 
\begin{align}
  BR (\tilde{T} \to W \, b) &\sim 0.5 \, , \nonumber \\
  BR (\tilde{T} \to Z \, t) &\sim 0.25 \, , \nonumber \\
  BR (\tilde{T} \to h \, t) &\sim 0.25 \, ,
  \label{M15decayBR}
\end{align}
while the branching ratios of the charge-2/3 top partners of $M4_{5}$
are given by 
\begin{align}
  BR (X_{2/3} \to Z \, t) \sim BR (T \to Z \, t)
  &\sim 0.5 \, , \nonumber \\ 
  BR (X_{2/3} \to h \, t) \sim BR (T \to h \, t)
  &\sim 0.5 \, . 
  \label{M45decayBR}
\end{align}

Besides the composite Higgs models, there are other models predicting
an $SU(2)$-singlet top partner, e.g.~Little Higgs models. A prime
example is the Littlest Higgs Model with T-parity (LHT)
\cite{ArkaniHamed:2002qy,Cheng:2003ju,Cheng:2004yc}. Within the class
of strongly coupled UV-completions of the SM, Little Higgs models
represent an appealing realisation exploiting a natural separation
between the electroweak scale $v$ and the compositeness scale $\Lambda
= 4 \pi f$. This is realised through Collective Symmetry
Breaking. This mechanism forces the global symmetries, preventing the
generation of a Higgs mass term, to be broken by at least two 
operators: in this way, the Higgs mass-generating one-loop diagrams
are at most logarithmically divergent in $\Lambda$, while
quadratically divergent only at two-loop level. The realisation of
this mechanism requires the introduction of additional partner fields
in the scalar, vector boson and top sectors, in order to formulate
``collective'' couplings of the Higgs boson to the SM particles and
their respective partners.  

The Littlest Higgs model is based on a non-linear sigma model
describing the global spontaneous symmetry breaking at the scale $f
\sim \mathcal{O}(\text{TeV})$ 
\begin{equation}
  SU(5)/SO(5) \, .
\end{equation}
The mechanism for this symmetry breaking is not specified: the model
describes an effective theory valid up to the compositeness scale
$\Lambda = 4 \pi f$, where a strong sector as UV-completion is
assumed. For comprehensive reviews of the model details see
\cite{Han:2003wu,Hubisz:2004ft,Hubisz:2005tx,Chen:2006cs,Belyaev:2006jh,Blanke:2006eb}. In
here we just mention that, in addition to the SM particles, new
charged heavy vector bosons ($W^{\pm}_{\subh}$), a neutral heavy
vector boson ($Z_{\subh}$), a heavy photon ($A_{\subh}$), a top
partner ($T_{+}$) and a triplet of scalar heavy particles ($\Phi$) are
present: these heavy particles acquire masses of order $f$ from the
$SU(5)/SO(5)$ spontaneous breaking. Couplings of the Higgs to these
particles radiatively generate a potential for the Higgs boson,
triggering the electroweak symmetry breaking. 

As the original Littlest Higgs model suffers from severe
constraints from electroweak precision tests (EWPT), which could be
satisfied only in rather extreme regions of the parameter space
\cite{Csaki:2002qg,Kilian:2003xt,Reuter:2012sd}, these can be
evaded with the introduction of a custodial symmetry, ungauging some
of the symmetries~\cite{Kilian:2004pp,Kilian:2006eh}, or with the
introduction of a conserved discrete symmetry called 
T-parity~\cite{Cheng:2003ju,Cheng:2004yc}. Using the latter, the scale
$f$ can be as low as $\mathcal{O}(500 \, \text{GeV})$, resulting in a
rather low amount of fine-tuning to accommodate the observed Higgs
mass, together with not too suppressed production cross sections of new
particles~\cite{Hubisz:2005tx,Berger:2012ec,Asano:2006nr,Reuter:2012sd}. 

Recent studies including constraints from EWPT, Higgs observables and
results from direct searches for new particles, have set a lower bound
on the scale $f$ to be
\cite{Reuter:2012sd,Reuter:2013zja,Reuter:2013iya} 
\begin{align}
  \left( f_{\text{\tiny LHT, A}} \right)_{\text{\tiny
      EWPT+Higgs}} &\gtrsim \ 694 \ \textrm{GeV} \\ 
  \left( f_{\text{\tiny LHT, B}} \right)_{\text{\tiny
      EWPT+Higgs}} &\gtrsim \ 560 \ \textrm{GeV} \, , 
\end{align}
depending on the particular implementation of the down Yukawa
couplings. The latter translate into e.g.~a lower bound on the mass of
the top partner  
\begin{align}
  \left( M_{T_{+}} \right)_{\text{\tiny LHT, A}} &\gtrsim \ 975
  \ \textrm{GeV} \\ 
  \left( M_{T_{+}} \right)_{\text{\tiny LHT, B}} &\gtrsim \ 787
  \ \textrm{GeV} \, . 
\end{align}

Besides the (T-even) top partner $T_{+}$, which is introduced to
regularise the quadratic divergence of the Higgs mass from the SM top
loop, a consistent implementation of T-parity in the top sector
requires the introduction of a T-odd counterpart of the heavy top
partner, called $T_{-}$, and a T-odd partner of the (T-even) SM top,
called $t_{\subh}$. While the introduction of the former is specific
for the top sector, every SM fermion is instead required to possess a
T-odd partner, generically called mirror fermion. Both $T_{+}$ and
$T_{-}$ acquire a mass of order $f$ from a Yukawa-like Lagrangian, as
well as the SM top after electroweak symmetry breaking; on the other
hand, the mass generation for mirror fermions requires the
introduction of a Lagrangian involving couplings proportional to a new
free parameter $\kappa$. $R$ is a ratio of Yukawa couplings in the top
sector (for more details, cf.~e.g.~\cite{Reuter:2012sd}).

\begin{table}[!ht]
  \centering
  \small
  \begin{tabular}{l l r r} 
    \toprule[1pt]
    Particle & Decay & $\textrm{BR}_{\kappa=1.0}$ &
    $\textrm{BR}_{\kappa=0.4}$ \\ 
    \midrule[1pt]
    $d_\subh$ & $W_\subh^- \; u$ & $63\%$ & $0\%$ \\
    & $Z_\subh \; d$ & $31\%$ & $0\%$ \\
    & $A_\subh \; d$ & $6\%$ & $100\%$ \\
    \midrule 
    $u_\subh$ & $W_\subh^+ \; d$ & $61\%$ & $0\%$ \\
    & $Z_\subh \; u$ & $30\%$ & $0\%$ \\
    & $A_\subh \; u$ & $9\%$ & $100\%$ \\
    \midrule
    $T_+$ & $W^+ \; b$ & $46\%$ & $46\%$ \\
    & $Z \; t$ & $22\%$ & $22\%$ \\
    & $H \; t$ & $21\%$ & $21\%$ \\
    & $T_- \; A_\subh$ & $11\%$ & $11\%$ \\
    \midrule
    $T_-$ & $A_\subh \; t$ & $100\%$ & $100\%$ \\
    \bottomrule[1pt]
  \end{tabular}
  \caption{Overview of the decay modes with the corresponding
    branching ratios of the LHT new quarks, with reference
    values $f = 1 \, \textrm{TeV}$ and $R = 1.0$
    \cite{Reuter:2013zja,Reuter:2013iya}. We emphasise two
    possible scenarios, namely with the mirror quarks
    $q_{\subh}$ either lighter ($\kappa = 0.4$) or heavier
    ($\kappa=1.0$) than the gauge boson partners. The heavy
    leptons decay analogously to the heavy quarks and the decays
    involving generic up or down quarks have to be considered as
    summed over all flavours.} 
  \label{tab:lhtdecayBRs}
\end{table}

In table \ref{tab:lhtdecayBRs} we list an overview of decay modes and
branching ratios of the LHT new particles, with reference values $f =
1 \, \textrm{TeV}$ and $R = 1.0$. In particular, the LHT $T_{+}$ top
partner shares the 2:1:1 ratio for the decays into SM particles as in
eq.~\eqref{M15decayBR}, but allows for a further decay channel
involving the T-odd partner $T_{-}$ and the heavy photon $A_{\subh}$
with a non-negligible rate. 

The electroweak coupling of $T_{+}$ to the $W$ boson, which governs
its associated production with a b-quark, is given by
\cite{Blanke:2006eb} 
\begin{equation}
  \text{coeff} \left(W^{+} \, \bar{T}_{+ \, \subr} \, b_{\subl}
  \right) = \frac{g}{\sqrt{2}} \, \frac{R^{2}}{1+R^{2}} \,
  \frac{v}{f} + \mathcal{O} \left( \frac{v^{2}}{f^{2}} \right)
  \equiv \frac{g \, g^{\ast}}{\sqrt{2}} \, . 
  \label{LHTsingleprodCoupling}
\end{equation}
Note that we again put this into the same form as
eq.~\eqref{M15associatedcoupling}.

From this, it is clear that charge-2/3 vector-like top partners share
similar final state topologies, with different branching ratios and
single production couplings depending on the particular underlying
model. Therefore, when looking for possible dedicated searches for top
partners at the LHC, it is favourable to use simplified model
approaches, involving for example only the mass of the top partner and
its ``single production'' coupling as free parameters. We pursue this
approach for the rest of the paper.

%-- Subsection: Simplified model approach
\subsection{Simplified model approach}
Recently, a generic parametrisation of an effective Lagrangian for top
partners has been proposed in \cite{Buchkremer:2013bha}, where the
authors considered vector-like quarks embedded in different
representations of the weak $SU(2)$ group, with other minimal
assumptions regarding the structure of the couplings. In particular,
vector-like quarks which can mix and decay directly into SM quarks of
all generations are included. Particularly interesting for our
purposes is the case in which the top partner is an $SU(2)$ singlet,
with couplings only to the third generation of SM quarks. The
Lagrangian parametrising the possible top partner interactions reads
\cite{Buchkremer:2013bha} 
\begin{equation}
  \mathcal{L}_{\subt} \supset \frac{g^{\ast}}{\sqrt{2}} \left[
    \frac{g}{\sqrt{2}} \, \bar{T}_{\subl} \, W_{\mu}^{+}
    \gamma^{\mu} \, b_{\subl} + \frac{g}{2 c_{\subw}} \,
    \bar{T}_{\subl} \, Z_{\mu} \gamma^{\mu} \, t_{\subl} -
    \frac{M_{\subt}}{v} \, \bar{T}_{\subr} \, h \, t_{\subl} -
    \frac{m_{t}}{v} \, \bar{T}_{\subl} \, h \, t_{\subr} \right]
  + \, \text{h.c.} \, , 
  \label{TsingletVL}
\end{equation}
where $M_{\subt}$ is the top partner mass, and $g^{\ast}$ parametrises
the single production coupling in association with a b- or a
top-quark. In the limit of $M_{\subt} \gg m_{t}$, the width of the
top partner is
\begin{equation}
  \Gamma_{\subt} \simeq \frac{\left( g \, g^{\ast} \right)^{2}
    \, M_{\subt}^{3}}{64 \, \pi \, m_{\subw}^{2}} \left( 1 +
    \frac{1}{2} + \frac{1}{2} \right) \, , 
\end{equation}
where the three contributions in parentheses arise from the top partner
decays to $W$, $Z$ and Higgs, respectively. The different branching
ratios of $T$ are thus clearly the same as in eq.~\eqref{M15decayBR},
since we are describing effectively the same type of top partner as in
$M1_{5}$. 

For our proposed top partner search at the LHC we will exploit a
simplified-model approach, assuming the interactions described by the
Lagrangian of eq.~\eqref{TsingletVL}, where the only free parameters
will be the top partner mass $M_{\subt}$ and its ``single production''
coupling $g^{\ast}$. In this way, our results will be
straightforwardly mapped within the context of the $M1_{5}$ minimal
Composite Higgs model, namely by identifying as in
eq.~\eqref{M15associatedcoupling} 
\begin{equation}
  y = \frac{g \, g^{\ast}}{\sqrt{2}} \,
  \frac{M_{\tilde{\subt}}}{m_{\subw}} \qquad (M1_{5}) \, .  
  % \quad \Rightarrow \quad g^{\ast} = \sin \alpha
\end{equation}
For comparison, with $y = 1$ and $M_{\tilde{\subt}} = 1 \,
\textrm{TeV}$ one obtains $g^{\ast} \sim 0.17$. 

On the other hand, while an immediate map of $g^{\ast}$ to the LHT
parameters is straightforward from eq.~\eqref{LHTsingleprodCoupling},
namely with 
\begin{equation}
  g^{\ast} = \sqrt{2} \, \frac{R^{2}}{1+R^{2}} \, \frac{v}{f} +
  \mathcal{O} \left( \frac{v^{2}}{f^{2}} \right) \qquad
  (\text{LHT}) \, , 
\end{equation}
the Lagrangian of eq.~\eqref{TsingletVL} does not exactly reproduce
the $T_{+}$ phenomenology because of the absence of the $T_{+}
\to T_{-} \, A_{\subh}$ vertex in the simplified-model
approach. In particular, it should be kept in mind that the different
branching ratios of the top partner 
described by eq.~\eqref{TsingletVL} slightly overestimate the actual
branching ratios of the LHT $T_{+}$ partner. For comparison, fixing $R
= 1.0$ and $f = 1 \, \textrm{TeV}$ yields $g^{\ast} \sim 0.17$. 

Finally, by using the simplified-model approach, we also underestimate
the branching ratios of the charge-2/3 top partners within the
$M4_{5}$ model, given in eq.~\eqref{M45decayBR}: our results will be
conservative in this case. 

%-- Subsection: Tagging the boosted regime
\subsection{Tagging the boosted regime}
Let us now focus on the kinematics of a possible top partner
decay. For masses much heavier than the top quark, the top partner
decay products are produced with large spatial separation
(back-to-back decay). Furthermore, for large center-of-mass
energies, these primary top partner decay products are
necessarily boosted, namely with transverse momentum $p_{\subt}$ which
considerably exceeds their rest mass: this means that the subsequent
decay products are highly collimated in one area of the detector. As a
rule of thumb, the decay products of a highly boosted particle of mass
$m$ and transverse momentum $p_{\subt} \gg m$ are collimated within a
cone of radius 
\begin{equation}
  \Delta R \sim 2 \, \frac{m}{p_{\subt}} \, ,
  \label{deltaRboostedtop}
\end{equation}
such that e.g.~the hadronic decays of a boosted SM top with $p_{\subt}
\sim 250 \, \textrm{GeV}$ are collimated within a detector region of
radius $\Delta R \lesssim 1.4$. 

In this kinematical regime, conventional reconstruction algorithms
that rely on a jet-to-parton assignment are often
not feasible. Crucial ingredients for high center-of-mass
searches involving massive particles are the so-called substructure
methods \cite{Butterworth:2008iy,Plehn:2010st}, to identify the top
partner decay products within large ``fat'' jets. Generically,
focusing on hadronic decays of boosted objects, these substructure
methods first reconstruct jets with a much larger radius parameter, in
order to capture the energy of the complete hadronic decay in a single
jet; then use method-dependent discriminating variables to analyse
the internal structure of the fat jets, to separate boosted objects
from the large QCD background.  

Jet-substructure methods which are dedicated to the identification of
possible boosted tops are generically called
\emph{top}-\emph{taggers}. In particular, top tagging techniques are
crucial not only to reduce the huge SM QCD and $t \bar{t}$
backgrounds, exploiting the particular kinematical feature of the
boosted decay products, but also to avoid combinatorics in the
reconstruction of the top four momentum from high multiplicity
final-state jets. In this way, fully-hadronic top decays with a
larger branching ratio compared to leptonic final states, can be
systematically exploited for searches involving top partners. A review
on top-taggers can be found e.g.~in \cite{Plehn:2011tg}.  

It turns out, see e.g.~refs.~\cite{CMS:2014fya,Yang:2014usa}, that the
Heidelberg-Eugene-Paris top-tagger \cite{Plehn:2010st}
(``HEPTopTagger'') can have a relatively better performance compared
to other algorithms, especially for moderately boosted tops. For this
reason, in our analysis we will adopt the HEPTopTagger to tag boosted
top quarks in the considered signal events.

%%%-- Section: Setup of the analysis --%%%
\section{Setup of the analysis}
\label{sec:ttagsetup}

%-- Subsection: Event generation
\subsection{Event generation}
\begin{figure}[!ht]
  \centering
  \includegraphics[width=.7\textwidth]{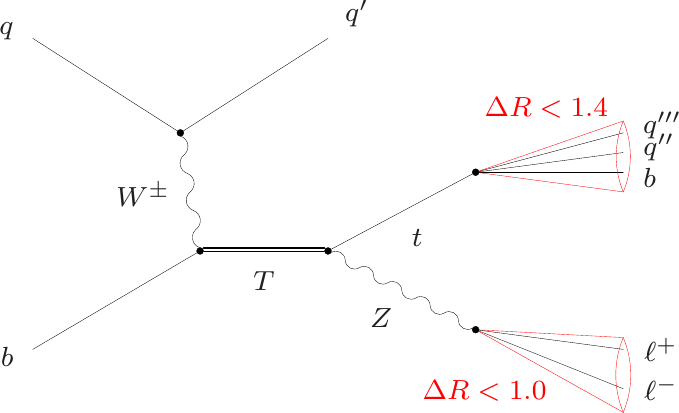}
  \caption{Single production of a heavy top partner $T$ with
    subsequent decay into $tZ$. The boosted decay products of the
    latter are collected inside cones of $\Delta R < 1.4$ and $\Delta
    R < 1.0$, respectively.}
  \label{fig:proc}
\end{figure}

As mentioned in section \ref{sec:toptaggingintro}, we investigate
processes involving a charge-2/3 vector-like top partner $T$,
inclusively pair and associated produced, with subsequent decay 
\begin{equation}
  T \to t \, Z \to \left( q \, q^{\prime} \, b
  \right) \, \left( \ell^{+} \ell^{-} \right) \, . 
  \label{TZtdecay}
\end{equation}
The process is depicted in figure \ref{fig:proc} together with our
conditions on the cones of the boosted objects to be defined below. 
We study a possible search strategy optimised for the LHC with
center-of-mass energy of $\sqrt{s} = 13 \, \textrm{TeV}$ and
integrated luminosity of $300 \, \textrm{fb}^{-1}$. The clean final
state and the absence of missing transverse energy makes this channel
promising for a possible mass reconstruction of the top partner, even
if the possible SM backgrounds are rather huge. 

Signal and background events have been simulated using
MadGraph5 v2.1~\cite{Alwall:2014hca}, and
Pythia 8.183 \cite{Sjostrand:2007gs} for parton-shower and
fragmentation, and further analysed via Delphes 3.1
\cite{deFavereau:2013fsa} for a fast detector simulation following the
specifications which we are going to detail in the following. All
cross sections have been checked with 
WHIZARD v2.2~\cite{Kilian:2007gr,Moretti:2001zz,Kilian:2011ka}. In
particular, an anti-$k_{t}$ jet clustering algorithm with radius
parameter of $R = 0.4$ is used to reconstruct jets, which in the
following we will call \emph{slim jets}. The same Pythia output is
simultaneously analysed through FastJet 3.0.6
\cite{Cacciari:2011ma,Cacciari:2005hq} in order to cluster the
hadronic activity using the Cambridge-Aachen algorithm with larger
radius parameter of $R = 1.5$, reconstructing jets which in the
following we will identify as \emph{fat jets}. 

The model file generating signal events according to the Lagrangian of
eq.~\eqref{TsingletVL} \cite{Buchkremer:2013bha}, can be found in the
dedicated FeynRules model database webpage (``Singlet T Model VLQ'')
\cite{Christensen:2008py,Degrande:2011ua,Christensen:2010wz}. The
corresponding free parameters are the top partner mass $M_{\subt}$,
the coupling $g^{\ast}$ which governs the top partner single
production involving a t-channel $W$, and the rate $R_{\subl}$ of $T$
decays into light quarks. We fix $R_{\subl} = 0$ in order to force $T$
to decay only to third generation SM quarks. For our analysis we
consider values in the range 
\begin{equation}
  M_{\subt} \in \left[ 850, 1450 \right] \, \textrm{GeV} \, ,
  \qquad g^{\ast} \in \left[ 0.05, 0.5 \right] \, . 
  \label{parameterrange} 
\end{equation}

In particular, our signal processes consist of pair and associated
production of a charge-2/3 vector-like top partner $T$, with
subsequent decay as in eq.~\eqref{TZtdecay}: in the case of pair
production we consider the inclusive decay of the second top partner
according to the branching ratios reported in
eq.~\eqref{M15decayBR}. The LO signal cross section is calculated via
MG5, depending on the particular choice of the free parameters
which were consistently updated, together with the top partner width,
before the event generation. We further rescale the signal cross
section with a K-factor which we evaluate using Hathor 2.0
\cite{Aliev:2010zk,Kant:2014oha}. In particular, we calculate the
K-factors for both top pair (NNLO) and single productions (NLO) for
different values of the top mass in the range \eqref{parameterrange},
eventually choosing a minimal and hence conservative value of $K = 1.14$. 

The main SM background processes turn out to be $Z + \, \text{jets}$,
associated $Z$ production with a pair of top quarks ($t\bar{t} \, Z +
\, \text{jets}$), plus subleading contributions from associated $Z$
production with single top ($t/\bar{t} \, Z + \, \text{jets}$). All
other potentially dangerous contributions like $t \bar{t} + \,
\text{jets}$, $t \bar{t} \, W^{\pm} + \, \text{jets}$ and
$\gamma^{\ast} \to \ell^{+} \ell^{-} + \, \text{jets}$ turn
out to be negligible by requiring exactly two opposite charge and same
flavour leptons in the final state with invariant mass satisfying
$|m_{\ell^{+} \ell^{-}} - m_{\subz}| < 10 \, \textrm{GeV}$. Furthermore, 
the large $W^{\pm}Z + \, \text{jets}$ background becomes also negligible  
due to the smaller boost of the $Z$ boson compared to the signal and the 
backgrounds involving the top quark, and by exploiting b-~and top-tagging.

Large samples of background events are generated using MG5,
requiring up to three, two or one additional hard jets at matrix
element level for $Z + \, \text{jets}$, $t/\bar{t} \, Z + \,
\text{jets}$ and $t\bar{t} \, Z + \, \text{jets}$ processes,
respectively. To avoid double counting of jets generated at matrix
element level and jets radiated during the parton showering process, a
CKKW-L merging procedure
\cite{Catani:2001cc,Lonnblad:2001iq,Lonnblad:2011xx} is exploited. In
particular we interface, for each background sample, the corresponding
parton level MG5 outputs with different multiplicities of
additional jets to Pythia 8.183 and its internally built-in routines
for the CKKW-L merging, accordingly setting the merging scale value
and the number of additional jets available from matrix element. This
procedure guarantees a correct prediction for the (merged) cross
section of the desired process. 

\begin{table}[!ht]
  \centering
  \begin{tabular}{l | c | c}
    \toprule[1pt] 
    bkg.~process & K-factor & Ref. \\ 
    \midrule[1pt]
    $Z + \, \text{jets}$ & 1.20  & \cite{Catani:2009sm} \\ 
    $t\bar{t} \, Z + \, \text{jets}$ & 1.30 &
    \cite{Kardos:2011na} \\ 
    $t \, Z + \, \text{jets}$ & 1.11 & \cite{Campbell:2013yla} \\
    $\bar{t} \, Z + \, \text{jets}$ & 1.09 &
    \cite{Campbell:2013yla} \\ 
    \bottomrule[1pt]
  \end{tabular}
  \caption{K-factors of the leading SM background processes for
    our analysis.} 
  \label{table:kfactors}
\end{table}%

We rescale the evaluated background cross sections with appropriate
K-factors from the corresponding publications, summarising the values
in table \ref{table:kfactors}. It should be noted that the inclusive
$t\bar{t} \, Z + \, \text{jets}$ K-factor as given in
\cite{Kardos:2011na} is $K = 1.39$: however, this value is reduced for
large top transverse momenta, as in our case. For this reason we
conservatively set $K = 1.30$ as in table \ref{table:kfactors}. 

%-- Subsection: Reconstruction of physics objects
\subsection{Reconstruction of physics objects}
Final state object reconstruction is performed mainly following the
specifications detailed in \cite{ATLAS:2012mn}. An electron candidate
is required to have a transverse momentum $p_{\subt}^{e} \geq 20 \,
\textrm{GeV}$ and $|\eta^{e}| < 2.47$. An isolation requirement is
further applied, namely the total $p_{\subt}$ of all charged particles
$q$ satisfying $p_{\subt}^{q} > 1.0 \, \textrm{GeV}$ and $\Delta R (e,
q) < 0.3$, should be less than 10$\%$ of $p_{\subt}^{e}$. A muon
candidate is required to satisfy $p_{\subt}^{\mu} \geq 10 \,
\textrm{GeV}$ and $|\eta^{\mu}| < 2.5$. The isolation for the muon
requires that the total $p_{\subt}$ of all charged particles $q$
satisfying $p_{\subt}^{q} > 1.0 \, \textrm{GeV}$ and $\Delta R (\mu,
q) < 0.4$, should be less than 6$\%$ of $p_{\subt}^{\mu}$. %The
                                %identification requirements are
                                %adjusted to be $90\%$ efficient for
                                %charged leptons from $W$ and $Z$
                                %decay. 

As mentioned before, slim jets are clustered from all final state
particles with $|\eta|<4.9$, except isolated leptons and neutrinos,
using the anti-$k_{t}$ algorithm with a radius parameter of $R = 0.4$
as implemented in Delphes 3.1. Only slim jets with $p_{\subt}^{j}
\geq 20 \, \textrm{GeV}$ are further considered. Slim jets are
possibly identified as b-jets through the built-in Delphes 3.1
dedicated routines: in particular, we set the probability to tag
b-jets (b-tag efficiency) to 70$\%$, together with a charm quark
misidentification probability of 10$\%$. Tagged b-jets are further
required to be reconstructed within $|\eta^{b}| < 2.5$. 

Fat jets are simultaneously clustered using FastJet 3.0.6 on the same
final state particles with $|\eta|<4.9$, except isolated leptons and
neutrinos, using the Cambridge-Aachen algorithm with radius parameter
of $R = 1.5$. Only fat jets with $p_{\subt}^{j} \geq 20 \,
\textrm{GeV}$ are further considered. 

%-- Subsection: Cutflow
\subsection{Cutflow}
Events are required to contain in the final state at least two leptons
with minimum transverse momentum $p_{\subt}^{\ell} > 25 \,
\textrm{GeV}$. Among all possible pairs of leptons, we require at
least one pair to consist of opposite charge and same flavour leptons
matching the invariant mass of the $Z$ boson, namely such that the
lepton-pair invariant mass $m_{\ell^{+} \ell^{-}}$ satisfies 
\begin{equation}
  |m_{\ell^{+} \ell^{-}} - m_{\subz}| < 10 \, \textrm{GeV} \, .
\end{equation}
We further require that, for at least one pair, the separation $\Delta
R = \sqrt{\Delta \phi^{2} + \Delta \eta^{2}}$ between the two
candidate leptons reconstructing the $Z$ mass should satisfy 
\begin{equation}
  \Delta R (\ell^{+}, \ell^{-}) < \Delta R (\ell^{+},
  \ell^{-})_{\text{max}} = 1.0 \, . 
  \label{deltaRcut}
\end{equation}
If more than one pair of leptons satisfies the previous requirements,
we select the pair with invariant mass closest to the $Z$ boson
mass. This pair of leptons allows us to fully reconstruct the
four-momentum of the candidate $Z$ boson. 

The cut of eq.~\eqref{deltaRcut} is particularly effective to suppress
SM backgrounds containing a $Z$ boson, since it captures the expected
boosted kinematics of the $Z$ boson from the top partner
decay. According to eq.~\eqref{deltaRboostedtop}, we expect indeed
highly collimated decay products from a boosted $Z$. On the other
hand, SM processes do not provide a large transverse boost to the $Z$
boson, guaranteeing a good discrimination power to
eq.~\eqref{deltaRcut}. 

\begin{figure}[!ht]
  \begin{center}
    \includegraphics[width=0.7\textwidth]{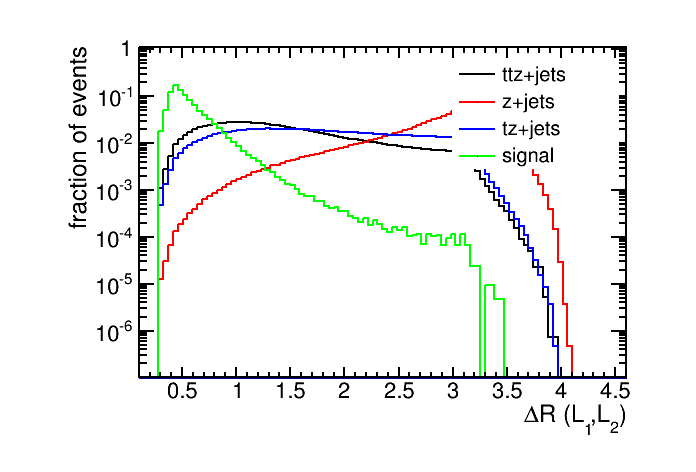}
    \caption{Distribution of the $\Delta R$ variable
      evaluated among candidate leptons reconstructing the
      $Z$ boson for different processes. The signal
      process assumes $M_{\subt} = 1 \, \textrm{TeV}$ and
      $g^{\ast} = 0.1$.} 
    \label{fig:DeltaRLL}
  \end{center}
\end{figure}

We show in figure~\ref{fig:DeltaRLL} the distribution of the variable
$\Delta R$ evaluated among candidate leptons reconstructing the $Z$
boson, for the different background and signal processes: a peak at
smaller values of $\Delta R$ is clearly visible for signal
events. Note that the signal events used for all distribution plots
shown in this section correspond to the benchmark point $M_{\subt} = 1
\, \textrm{TeV}$ and $g^{\ast} = 0.1$. 

\begin{figure}[!ht]
  \begin{center}
    \includegraphics[width=0.7\textwidth]{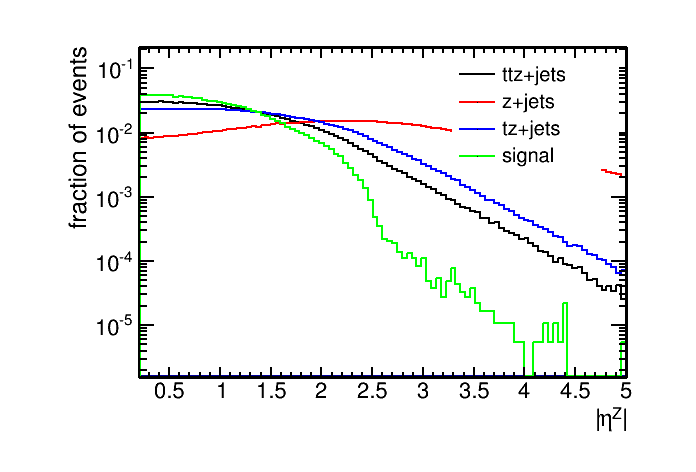} 
    \caption{Distribution of the psudorapidity
      $|\eta^{\subz}|$ of the reconstructed candidate $Z$
      boson for different processes. The signal process
      assumes $M_{\subt} = 1 \, \textrm{TeV}$ and
      $g^{\ast} = 0.1$.} 
    \label{fig:etaZ}
  \end{center}
\end{figure}

Further kinematic constraints are imposed on the candidate $Z$ boson,
again to exploit the boosted properties of the considered signal. In
particular, we require a large transverse momentum of the candidate
$Z$, namely 
\begin{equation}
  p_{\subt}^{\subz} > p_{\subt, \, \text{min}}^{\subz} = 225 \,
  \textrm{GeV} \, , 
\end{equation}
as well as requiring that the $Z$ should be produced in the central
region of the detector, namely with 
\begin{equation}
  |\eta^{\subz}| < |\eta^{\subz}|_{\text{max}} = 2.3 \, .
  \label{etaZcut}
\end{equation}
The requirement of eq.~\eqref{etaZcut} is useful in rejecting e.g.~the
SM $Z + \, \text{jets}$ background, the latter being mostly produced
via a Drell-Yan process with the initial quarks yielding a forward
boost to the produced $Z$ boson, as can be seen in figure
\ref{fig:etaZ}. 

\begin{figure}[!ht]
  \begin{center}
    \includegraphics[width=0.7\textwidth]{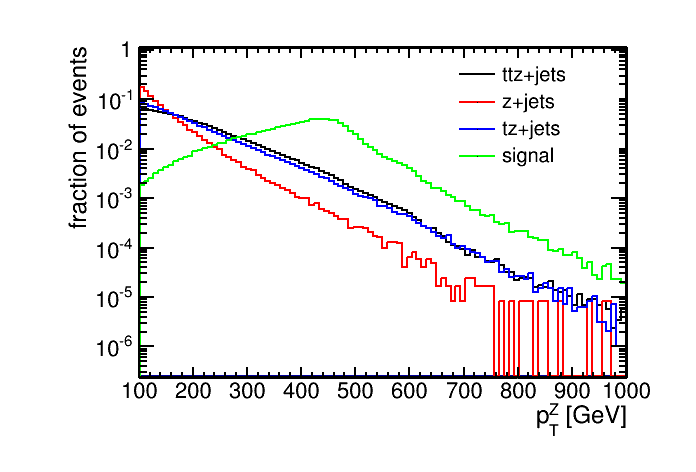}
    \caption{Distribution of the transverse momentum
      $p_{\subt}^{\subz}$ of the reconstructed candidate
      $Z$ boson for different processes. The signal
      process assumes $M_{\subt} = 1 \, \textrm{TeV}$ and
      $g^{\ast} = 0.1$.} 
    \label{fig:ptZ}
  \end{center}
\end{figure}

In figure \ref{fig:ptZ} we show the distribution of the transverse
momentum of reconstructed $Z$ boson candidates as described in the
text. Larger transverse momenta are observed for the (boosted) $Z$
from the signal process. 

\begin{figure}[!ht]
  \begin{center}
    \includegraphics[width=0.7\textwidth]{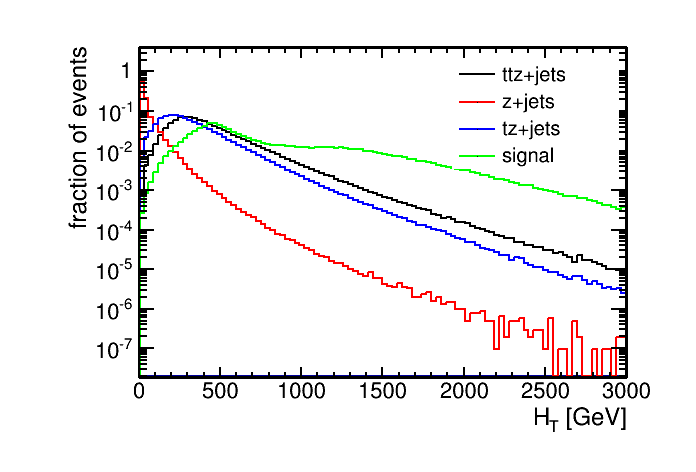} 
    \caption{Distribution of the scalar sum of the
      transverse momenta $H_{\subt}$ of the clustered
      \emph{slim jets} for different processes. The
      signal process assumes $M_{\subt} = 1 \,
      \textrm{TeV}$ and $g^{\ast} = 0.1$.} 
    \label{fig:HT}
  \end{center}
\end{figure}

In the next step, the hadronic activity is considered for additional
selection cuts. In order to account for the large boost of the top
quark, we expect the final state jets to possess a large amount of
transverse momentum. Therefore, we evaluate the $H_{\subt}$ variable,
namely the scalar sum of the transverse momenta of the reconstructed
slim jets with $p_{\subt}^{j} > 30 \, \textrm{GeV}$ and within
$|\eta^{j}| < 3.0$, requiring each event to satisfy 
\begin{equation}
  H_{\subt} > H_{\subt, \, \text{min}} = 400 \, \textrm{GeV} \, .
  \label{htcut}
\end{equation}

In figure \ref{fig:HT} we show the $H_{\subt}$ distribution for the
different considered processes. The signal distribution has a
considerable tail for larger values of $H_{\subt}$ compared to
background events, confirming the good discrimination power of
eq.~\eqref{htcut}. It is also worth noticing that the $H_{\subt}$
distribution for the signal in figure \ref{fig:HT} displays two
different visible peaks, at $\mathcal{O}(500 \, \textrm{GeV})$ and at
$\mathcal{O}(1.3 \, \textrm{TeV})$: these correspond to the top
partner single and pair production components of the signal,
respectively. 

Among the reconstructed final state slim jets, we further require
the presence of at least one tagged b-jet with 
\begin{equation}
  p_{\subt}^{b} > p_{\subt, \, \text{min}}^{b} = 40 \, \textrm{GeV} \, .
\end{equation}

We then turn our attention to the reconstructed fat jets in the final
state: our aim is to identify one reconstructed fat jet as our top
candidate. At least one fat jet is required to be reconstructed among
final state particles, satisfying the definition of fat jets given
before, and with an additional requirement on its transverse momentum
being 
\begin{equation}
  p_{\subt}^{J} > p_{\subt, \, \text{min}}^{J} = 200 \, \textrm{GeV} \, .
\end{equation}

Most importantly, we require at least one fat jet to be
HEPTop-tagged: the presence of a boosted SM top from the decay of a
heavier resonance is indeed one of the main features of the signal. As
mentioned in section \ref{sec:toptaggingintro}, top tagging is crucial
not only as a discriminant against SM backgrounds, but also to
effectively deal with the combinatorics in the top reconstruction from
high multiplicity final state jets. If more than one fat jet is
identified as a (boosted) top jet via the HEPTopTagger algorithm, we
identify our candidate top as the fat jet mostly back-to-back with
respect to the previously reconstructed candidate $Z$ direction, as we
would expect from the signal topology. 

\begin{figure}[!ht]
  \begin{center}
    \includegraphics[width=0.7\textwidth]{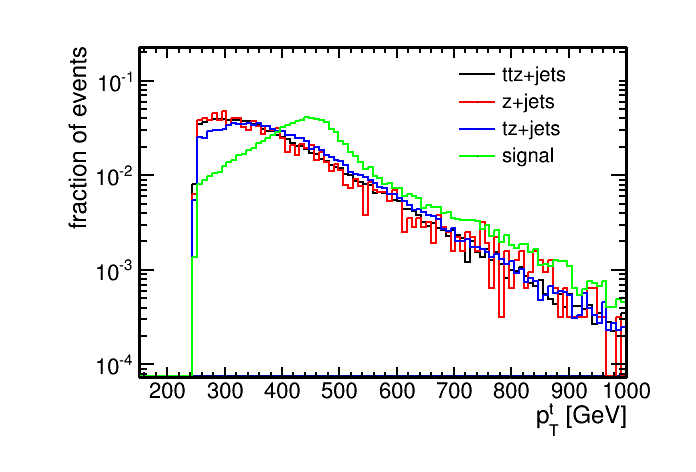}
    \caption{Distribution of the transverse momentum
      $p_{\subt}^{t}$ of the reconstructed candidate top
      for different processes. The signal process assumes
      $M_{\subt} = 1 \, \textrm{TeV}$ and $g^{\ast} =
      0.1$.} 
    \label{fig:ptT}
  \end{center}
\end{figure}

To account for its boosted kinematics, we require that the transverse
momentum of the candidate top should satisfy the cut 
\begin{equation}
  p_{\subt}^{t} > p_{\subt, \, \text{min}}^{t} = 250 \, \textrm{GeV} \, .
  \label{ptTcut}
\end{equation}
The $p_{\subt}^{t}$ distribution of signal and background processes,
after applying the cut of eq.~\eqref{ptTcut}, is shown in figure
\ref{fig:ptT}: a large fraction of signal events is observed for
higher values of $p_{\subt}^{t}$. 

Finally, to ensure that at least one of the tagged b-jets is
originating from the candidate top, and not from additional radiation
or as decay product of another involved particle, we require that the
spatial separation between the candidate top and at least one of the
slim jets tagged as b-jet should satisfy 
\begin{equation}
  \Delta R (t, b) < \Delta R (t, b)_{\text{max}} = 0.8 \, .
\end{equation}
In other words, this cut ensures that at least one (slim) b-jet
lies within the decay-cone of the candidate (fat jet) top. 

\begin{table}[!ht]
  \centering
  \begin{tabular}{l | r}
    \toprule[1pt]
    \multicolumn{1}{c}{} & selection cuts \\
    \midrule[1pt]
    \multirow{5}{*}{reconstructed $Z$} & $n_{\ell^{+}
      \ell^{-}} \geq 1$ \\ 
    & $|m_{\ell^{+} \ell^{-}} - m_{\subz}| < 10 \, \textrm{GeV}$ \\
    & $\Delta R (\ell^{+}, \ell^{-}) < 1.0$ \\
    & $p_{\subt}^{\subz} > 225 \, \textrm{GeV}$ \\
    & $|\eta^{\subz}| < 2.3$ \\
    \midrule
    \multirow{2}{*}{slim jets} & $H_{\subt} > 400 \,
    \textrm{GeV}$ \\ 
    & $n_{b} \geq 1, \, p_{\subt}^{b} > 40 \, \textrm{GeV}$ \\
    \midrule
    \multirow{4}{*}{fat jets} & $n_{J} \geq 1, \,
    p_{\subt}^{J} > 200 \, \textrm{GeV}$ \\ 
    & HEPTop $n_{t} \geq 1$ \\
    & $p_{\subt}^{t} > 250 \, \textrm{GeV}$ \\
    & $\Delta R (t, b) < 0.8$ \\
    \bottomrule[1pt]
  \end{tabular}
  \caption{Summary of the selection cuts of the proposed
    analysis, sorted per type of reconstructed object on which
    the cut is applied.} 
  \label{table:cutflow}
\end{table}%

To summarise the applied cuts, in table \ref{table:cutflow} we
categorise them according to the reconstructed object on which they
are applied. It should be noted that the actual values of $\Delta R
(\ell^{+}, \ell^{-})_{\text{max}}$, $p_{\subt, \,
  \text{min}}^{\subz}$, $|\eta^{\subz}|_{\text{max}}$, $H_{\subt, \,
  \text{min}}$, $p_{\subt, \, \text{min}}^{b}$ are identified using an
optimisation procedure: in particular, we scan the aforementioned cut
values within appropriate ranges and evaluate the corresponding signal
and background efficiencies for each possible configuration, obtaining
a signal over background (S/B) map as a function of the cut values. We
then identify the optimal cut configuration yielding the highest S/B
ratio, assuming $M_{\subt} = 1 \, \textrm{TeV}$ and $g^{\ast} = 0.1$
for the signal, and making sure that the total number of events after
applying the cuts would remain reasonably large for $300 \,
\textrm{fb}^{-1}$ of integrated luminosity.

\begin{table}[!ht]
  \centering
  \begin{tabular}{l | c | c | c}
    \toprule[1pt]
    selection cut & signal & $t\bar{t} \, Z + \,
    \text{jets}$ & $t \, Z + \, \text{jets}$ \\ 
    \midrule
    $n_{\ell^{+} \ell^{-}}, \, m_{\ell^{+} \ell^{-}}, \,
    \Delta R (\ell^{+}, \ell^{-})$ & $40.5 \%$ & $9.0 \%$
    & $4.9 \%$ \\ 
    $p_{\subt}^{\subz} > p_{\subt, \, \text{min}}^{\subz}$
    & $96 \%$ & $69 \%$ & $68 \%$ \\ 
    $|\eta^{\subz}| < |\eta^{\subz}|_{\text{max}}$ & $99
    \%$ & $99 \%$ & $99 \%$ \\ 
    $H_{\subt} > H_{\subt, \, \text{min}}$ & $80 \%$ & $64
    \%$ & $61 \%$ \\ 
    $n_{b} \geq 1, \, p_{\subt}^{b} > p_{\subt, \,
      \text{min}}^{b}$ & $77 \%$ & $72 \%$ & $55 \%$ \\ 
    $n_{J} \geq 1, \, p_{\subt}^{J} > p_{\subt, \,
      \text{min}}^{J}$ & $99 \%$ & $96 \%$ & $97 \%$ \\ 
    HEPTop $n_{t} \geq 1$ & $40 \%$ & $36 \%$ & $29 \%$ \\
    $p_{\subt}^{t} > p_{\subt, \, \text{min}}^{t}$ & $95
    \%$ & $82 \%$ & $85 \%$ \\ 
    $\Delta R (t, b) < \Delta R (t, b)_{\text{max}}$ & $80
    \%$ & $67 \%$ & $79 \%$ \\ 
    \midrule[1pt]
    final efficiency & $7.4 \%$ & $0.5 \%$ & $0.2 \%$ \\
    \midrule[1pt]
    production cross section [pb] & $1.2 \cdot 10^{-3}$ &
    $3.0 \cdot 10^{-2}$ & $1.9 \cdot 10^{-2}$ \\ 
    \bottomrule[1pt]
  \end{tabular}
  \caption{Efficiencies of the selection cuts evaluated on the
    considered processes. In particular, the signal events have
    been generated for the benchmark scenario $M_{\subt} = 1 \,
    \textrm{TeV}$, $g^{\ast} = 0.1$.} 
  \label{table:cutfloweff}
\end{table}%

In table \ref{table:cutfloweff} we collect the resulting
efficiencies evaluated on the different processes, together with the
corresponding production cross sections before the application of the
cuts. 

A final remark is devoted to possible pile-up effects, which we have
not explicitly included in our analysis. It is expected that at the
increased LHC center-of-mass energy runs and higher integrated
luminosity, an average of more than 50 interactions per proton-bunch
crossing will be observed. In particular, pile-up contamination might
shift mass distributions to higher values and broaden them. Since its
effect scales as the jet area, jets with larger cone area are more
susceptible to pile-up contamination. A dedicated pile-up
``mitigation'' strategy is beyond the scope of our analysis, also
because it would require a detailed detector information, but will
certainly have to be taken into account in a possible experimental
analysis.  

However, we expect our results to remain robust against pile-up
effects, since our analysis mostly relies on the identification of
leptons and exploits the HEPTopTagger to test the hadronic activity,
with an effective soft-radiation rejection already built-in through
a filtering procedure. In a recent publication \cite{Backovic:2014uma}
a thorough discussion has been presented of a possible search
strategy for top partners including an estimation of pile-up effects:
although being affected by pile-up contamination, the results of
their analysis are still consistent.

%%%-- Section: Results --%%%
\section{Results}
\label{sec:ttagresults}
The procedure detailed in section \ref{sec:ttagsetup} has a double
benefit, namely largely improving the S/B ratio on one hand, and
on the other hand uniquely determining the $4$-momenta of the
reconstructed top and $Z$ boson candidates satisfying the possible
kinematics of a top partner decay. 

\begin{figure}[!ht]
  \begin{center}
    \includegraphics[width=0.7\textwidth]{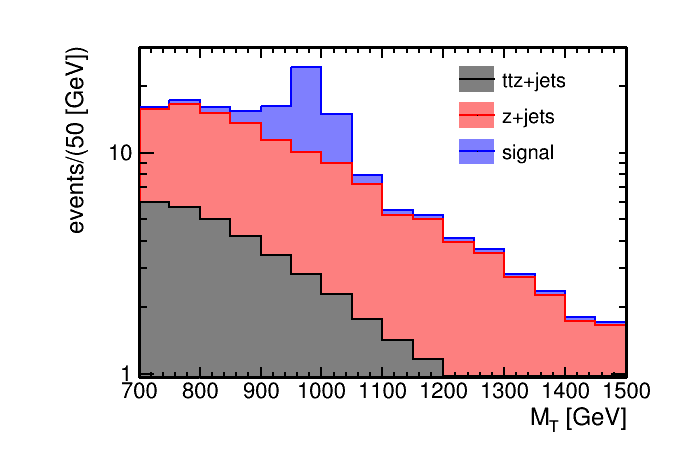}
    \caption{Stacked distribution plot of the invariant mass
      $M_{\subt}$ of the reconstructed top partner for different
      processes. All distributions have been rescaled with the visible
      cross section of the corresponding processes, times an
      integrated luminosity of $300 \, \textrm{fb}^{-1}$. The signal
      process assumes $M_{\subt} = 1 \, \textrm{TeV}$ and $g^{\ast} =
      0.1$. Other possible SM background processes are not shown in
      the plot since their contribution turned out to be negligible.} 
    \label{fig:thmass}
  \end{center}
\end{figure}

We finally plot the distribution of the invariant mass of the
($t$-$Z$) system, which we expect to peak at the invariant mass of
the on-shell top partner for the signal process, while described by a
smoothly descending distribution for the different backgrounds, since
the reconstructed top and $Z$ in the latter events do not originate
from an on-shell decay.  

We show the result in figure \ref{fig:thmass}, where we rescale the
different distributions with the visible cross section of the
corresponding processes, times an assumed integrated luminosity of
$300 \, \textrm{fb}^{-1}$. The different contributions are stacked in
the plot. In this way, figure \ref{fig:thmass} shows a realistic
amount of events which could be observed at the LHC with $\sqrt{s} =
13 \, \textrm{TeV}$ and $300 \, \textrm{fb}^{-1}$ of integrated
luminosity. For the signal we fixed $M_{\subt} = 1 \, \textrm{TeV}$
and $g^{\ast} = 0.1$. 

A peak in the bins around $M_{\subt} = 1 \, \textrm{TeV}$, fixing the
bin width to $50 \, \textrm{GeV}$, is clearly visible above the
background distribution, with up to 25 total events in the most
significant bin. The result of the analysis is therefore encouraging,
and we support the experimental collaborations to further analyse the
discussed channel: clearly, in a real experimental search the
background estimation would be more robust and precise, e.g.~via the
inclusion of reconstructed fake leptons. 

It is very important to estimate the significance of the signal peak
above the SM background, in order to consistently claim the evidence
for or the discovery of a top partner signal. In particular, the
hypothesis testing procedure is carried out using the public
BumpHunter code \cite{Choudalakis:2011qn}. This code operates on
datasets that are binned in some a-priori fixed set of bins: in our
case, the input datasets correspond to the total number of
signal$+$background and background-only events observed in
$M_{\subt}$-bins of $50 \, \textrm{GeV}$ as in figure
\ref{fig:thmass}. The BumpHunter scans the input-given data using a
window of varying width, and identifies the window with biggest excess
compared to the background: the dedicated test statistic is 
designed to be sensitive to local excesses of data\footnote{We setup
  the code to look for bumps in up to three consecutive bins, namely
  the possible mass resolution is at worst $\pm \, 75 \, \textrm{GeV}$
  around the central value.}. 

The same scanning procedure is further applied to pseudo-data sampled
from the expectation of the background input\footnote{In our case, we
  choose to model the background expectation by a Poisson distribution
  with the mean value distributed according to a Gamma
  distribution. The latter Gamma distribution is defined by fixing its
  mean value to the actual background bin value, and variance to the
  squared background bin error, as suggested in the BumpHunter
  manual. A total number of $10^{8}$ pseudo-experiments is generated
  accordingly.}, in order to reconstruct the ``expected'' distribution
of the test statistic. The $p$-value of the test is calculated, being
the probability that the test statistic will be equal to, or greater
than the test statistic obtained by comparing the actual data to the
background hypothesis. In other words, the $p$-value might be
interpreted as a false-discovery probability. When the distribution
of the test statistic is estimated using pseudo-experiments, as in
our case, then the $p$-value is calculated as a binomial success
probability.  

An equivalent formulation in terms of Gaussian significance is
straightforwardly obtained: it is common to claim that \emph{evidence}
for a new signal beyond the SM background is observed if the
$p$-value of the peak corresponds to at least $3.0 \sigma$ of
Gaussian significance, while it is common to claim a \emph{discovery}
if the $p$-value corresponds to at least $5.0\sigma$ of Gaussian
significance.  

By running the BumpHunter on the datasets summarised in figure
\ref{fig:thmass}, the most significant peak is observed in the
$[900,1050] \, \textrm{GeV}$ range, with an equivalent Gaussian
significance of $2.6^{+1.0}_{-0.9} \, \sigma$. The uncertainties on
the Gaussian significance of the peak are estimated by applying a
20$\%$ uncertainty on both the signal and background event yields,
which might account for up to 30$\%$ possible further non-statistical 
uncertainties which we have not taken into account. 

\begin{figure}[!ht]
  \begin{center}
    \includegraphics[width=0.7\textwidth]{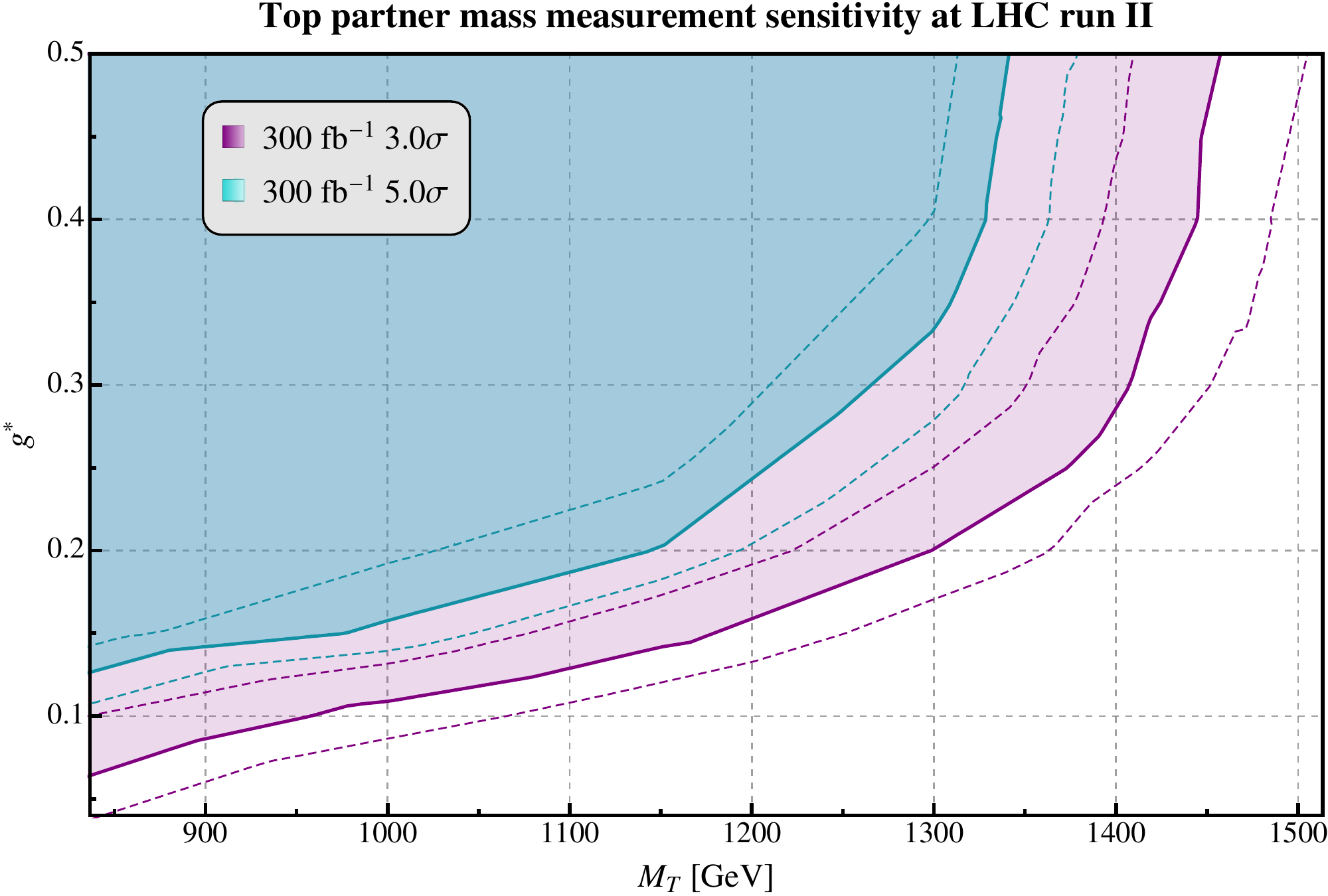}
    \caption{Parameter space regions of possible \emph{evidence}
      ($3.0\sigma$) or \emph{discovery} ($5.0\sigma$) of a top partner
      signal above the SM background, assuming the described analysis
      at the LHC with $\sqrt{s} = 13 \, \textrm{TeV}$ and $300 \,
      \textrm{fb}^{-1}$ of integrated luminosity. Also shown are bands
      representing the effect of a possible further non-statistical 30$\%$ 
      uncertainty on the visible cross section of the involved processes. If a
      signal peak is observed above the SM background, a possible mass
      measurement of the top partner invariant mass $M_{\subt}$ is
      possible with a mass resolution of at worst $\pm \, 75 \,
      \textrm{GeV}$ around the central value.} 
    \label{fig:thdiscovery}
  \end{center}
\end{figure}

Different hypotheses on the underlying BSM signal would alter the
shape of the signal distribution of figure \ref{fig:thmass}. However,
we expect that our analysis, although being optimised for the signal
values $M_{\subt} = 1 \, \textrm{TeV}$ and $g^{\ast} = 0.1$, should
still display a peak in the $M_{\subt}$ distribution even for
different choices of the free parameters. In particular, a higher
statistical significance of the peak might be achieved for different
signal hypotheses. For this reason, we generate a grid of signal
points for $M_{\subt} \in [850,1450] \, \textrm{GeV}$ in steps of $150
\, \textrm{GeV}$, and for $g^{\ast} \in [0.05,0.5]$ in steps of
$0.05$, and for each combination we evaluate the corresponding
significance of the peak, if observed.  

Our results are displayed in figure \ref{fig:thdiscovery}, where
regions of possible \emph{evidence} ($3.0\sigma$) or \emph{discovery}
($5.0\sigma$) of a top partner signal above the SM background are
identified, assuming a dedicated LHC analysis as discussed in the
text. Also shown are bands representing the effect of a possible total
30$\%$ uncertainty as discussed before. 
We observe that a large fraction of the considered 
parameter space might be probed using our proposed analysis; in
particular, the top partner mass might be measured via the described
BumpHunter procedure, with a mass resolution in our setup of at worst
$\pm \, 75 \, \textrm{GeV}$ around the central value. The mass
resolution might also be improved in a dedicated experimental setup.

From figure \ref{fig:thdiscovery} we see that the signal is within the range of 
possible evidence for top partner masses up to roughly $1450 \, \textrm{GeV}$ with 
$g^{\ast} \lesssim 0.5$, while being still sensitive to $g^{\ast}$ couplings down 
to $0.05$ at lower masses. The $g^{\ast} \to 0$ limit corresponds to 
the pair-production only component, being a QCD process independent 
on the electroweak coupling: one can observe that within our hypotheses 
and analysis setup, the single production component has to be necessarily 
non-vanishing to guarantee a possible discovery potential of the signal, 
since no discovery reach is obtained for values of $g^{\ast} \lesssim 0.05$. 
Analogously, for fixed top partner mass, the discovery potential 
increases with $g^{\ast}$, since the single production cross section grows as $|g^{\ast}|^{2}$.

We can now compare the discovery reach as presented in figure \ref{fig:thdiscovery}
with other existing studies in literature. In particular, we can first compare with the
results presented in \cite{Basso:2014apa}, where the ``trilepton'' decay 
channel $T \to t \, Z \to \left(b \, \ell \, \nu \right) \, \left( \ell^{+} \, \ell^{-} \right)$ has been scrutinised. 
In here, the authors considered a more general parameter space allowing mixing 
of the top partner with the other first two generations of quarks, 
namely letting the parameter $R_{\subl}$ to be non vanishing: this way, the 
production cross section of the top partner dramatically increases due to 
parton distribution enhancement, and the discovery reach becomes highly 
sensitive to $R_{\subl}$. The highest significance has been observed for $R_{\subl} \sim 1$, 
corresponding to $50\%$ mixing. The $R_{\subl} = 0$ case, as in our study, 
can be considered as the conservative case in which no flavour-changing
coupling is introduced. By comparing the discovery reach obtained in \cite{Basso:2014apa}
for $R_{\subl} = 0$ and $300 \, \textrm{fb}^{-1}$ of integrated luminosity, the trilepton 
and dilepton analyses show very similar results: the trilepton analysis of 
\cite{Basso:2014apa} extends the reach by $200 - 300 \, \textrm{GeV}$,
probing possible top partner masses up to roughly $1700 \, \textrm{GeV}$ with 
$g^{\ast} \lesssim 0.5$, while being still sensitive to $g^{\ast}$ couplings down to $0.1$.
Our dilepton search is instead more sensitive to lower values of the $g^{\ast}$
coupling, namely down to $g^{\ast} \sim 0.05$ for top partner masses of 
$850 \, \textrm{GeV}$. This is mainly due to the different b-jet cut requirement: while
in the trilepton analysis exactly one b-jet is required to be identified,
in our analysis we allow for the identification of more than one b-jet in the final state,
being thus more efficient in tagging the pair production component of the signal.

Although not immediate due to the different parameter space definitions, we can also draw
a comparison with other complementary studies for searches at the LHC run II involving 
a singlet top partner. In particular, in \cite{Endo:2014bsa} the authors 
show that a mass reconstruction is possible within the $T \to t \, h$ decay 
channel with $100 \, \textrm{fb}^{-1}$ of integrated luminosity at 
$\sqrt{s} = 14 \, \textrm{TeV}$, proposing a search strategy optimised for two 
top partner mass points, namely $m_{\subt} = 800, \, 900 \, \textrm{GeV}$,
and assuming $\textrm{BR}(T \to t \, h) = 1.0$. Furthermore, in 
\cite{Matsedonskyi:2014mna} the authors project at $\sqrt{s} = 13 \, \textrm{TeV}$ and 
$100 \, \textrm{fb}^{-1}$ of integrated luminosity the exclusion potential of the
analysis first presented in \cite{Ortiz:2014iza}, tailored for the leptonic 
$T \to W \, b$ decay channel with $\textrm{BR}(T \to W \, b) = 0.5$, obtaining
an exclusion reach up to $2.0 \, \textrm{TeV}$ for single production if 
$c_{\subl}^{\text{\tiny Wb}} \gtrsim 0.4$. Analogously, in \cite{Gripaios:2014pqa} 
the authors design a dedicated search strategy for the leptonic $T \to W \, b$ decay channel, 
obtaining an expected exclusion reach for masses up to $1.0 \, \textrm{TeV}$, including 
both pair and single production, with $\sqrt{s} = 14 \, \textrm{TeV}$ and $30 \, \textrm{fb}^{-1}$
of integrated luminosity. Our analysis is thus competitive with the results of existing 
literature, and represents a viable and complementary candidate to pursue the
search and mass measurement of a possible singlet top partner.

%%%-- Conclusion --%%%
\section{Conclusion}
\label{sec:conclusion}

In this paper we have investigated the search for new vector-like
heavy third-generation quarks, particularly top-like quarks in their
decay channel into a top quark and a $Z$ boson. Though this
neutral-current decay channel has not been thoroughly investigated yet
compared to the corresponding charged-current process into $Wb$ or the
decay into $th$, we believe that it is nevertheless worthwhile to look
into it: firstly, it offers another independent search channel, and
secondly the absence of missing transverse energy in the final
state allows for a complete mass determination of the heavy top
state. In order to be able to separate the fully hadronic top mode from
the huge SM backgrounds, we applied the techniques of boosted objects
and jet substructure to this channel. 

Such heavy vector-like top partners appear in many different BSM
models like models of (partial) compositeness, Little Higgs models,
extra-dimensional models etc. In order to be as model-independent as
possible we exploited a simplified model with only two free
parameters, the heavy top mass and an electroweak coupling constant.
We took both single and pair production of the heavy top quarks
into account, where generally single production is the less
phase-space constrained. The main SM backgrounds to these processes,
$Z$ + jets, $tZ$ + jets and $t\bar tZ$ + jets have been taken into account
using known NLO K-factors. The boost of the leptonically decaying $Z$
boson helps to suppress Drell-Yan backgrounds, while the signal is
discriminated by the fat jet characteristics of the collimated decaying
top quark. 

To determine the sensitivity of the upcoming run II of LHC to such
possible new states in this channel, we used the HepTopTagger to
discriminate fat top quark jets from SM backgrounds on simulated
events that have been merged with parton-shower generated QCD ISR and
FSR jets. Afterwards, the fast detector simulation from Delphes has
been used to assess efficiencies and uncertainties from the cut-flow
and the taggings. We briefly discussed possible pile-up contamination
and further non-statistical uncertainties. 

As a final result we gained the $3\sigma$ evidence reach as well as
the $5\sigma$ discovery potential of LHC run II in the parameter plane
of the two variables heavy top mass and effective coupling. This shows
that the discovery potential reaches up to roughly $1400 \, \textrm{GeV}$ for the
heavy top quark mass in regions of a still reliable heavy top quark
coupling. 

We encourage the experimental collaborations to look into this channel
as a possible discovery channel as well as a means to get direct
access to the mass of the heavy top with a final uncertainty of $75 \, \textrm{GeV}$
or better.

%%%-- Appendix--%%%
\appendix

%%%-- Acknowledgements --%%%
\acknowledgments
The authors of this paper are grateful for useful discussions with
Maikel de Vries, Lorenzo Basso, Stefan Prestel, Fabian Bach, Diptimoy Ghosh, Piero Ferrarese. 
MT has been partially supported by the Deutsche Forschungsgemeinschaft within
the Collaborative Research Center SFB 676 "Particles, Strings, Early
Universe".

%%%-- Bibliography --%%%
\bibliographystyle{JHEP}
\bibliography{top_partner.bib}

%%%-- End Document --%%%
\end{document}